# Energy Efficiency of Network-Coding Enabled Mobile Small Cells

Georgios P. Koudouridis[†], Henrik Lundqvist[†], Hong Li[‡], Xavier Gelabert[†]

[†]Huawei Technologies Sweden AB, Stockholm, Sweden
[‡]Huawei Technologies Co. Ltd., Shanghai, China
{george.koudouridis, henrik.lundqvist, macro.lihong, xavier.gelabert}@huawei.com

*Abstract*— Energy efficiency becomes increasingly important due to the limited battery capacity in wireless devices while at the same time user throughput requirements are relentlessly increasing. In this paper, we study an energy efficient cooperation scheme which employs network coding to enhance the energy efficiency for mobile devices. Herein we propose that the mobile devices are clustered into mobile small cells with one of the mobile devices acting as a group head with basic transceiver, coding and relaying functionalities. Group heads coordinate the transmissions from the mobile devices in the mobile small cell to the network's base stations. The objective function of the cooperative scheme is to minimize mobile devices' energy consumption subject to a certain bit error probability. The proposed network-coding based scheme has been evaluated by means of numerical simulations and compared to both a conventional direct transmit scheme, with no cooperation groups, and a cooperative relaying scheme. Results show that, with network-coded cooperation, energy efficiency may significantly increase provided the density of base stations and mobile devices is below a certain value. Above this value none of the compared cooperation schemes may improve energy efficiency, but rather power consumption is reduced only when mobile devices transmit via base stations in their close proximity.

## I. Introduction

Today's cellular networks are undergoing a major shift in their deployment and optimization due to an increasing number of connected devices and an unpreceded demand for wireless data traffic. This increasing density in traffic demand calls for further improvements in the efficiency of spatial reuse of spectrum resources [1]. Improving spectral efficiency requires a denser deployment of the radio access infrastructure and capacity-preserving transmission techniques [2].

Densification of a radio access network (RAN) implies the deployment of a large number of low-cost access nodes such as femto/pico base stations (BSs), fixed/mobile relays, distributed antennae, and radio heads (RHs). These may be connected to aggregation nodes by wireless links, which allows for the deployment of on-demand small cells, which are less involved than a full-functioning BSs, requiring mainly transceiver functionalities. The deployment of mobile devices to act as mobile small cells can further improve transmission capacity by employing cooperative schemes where user devices within the mobile small cell assist each other to cooperatively relay traffic.

A common concern with such solutions is the energy consumption of the mobile devices, which suffer from limited battery life time. Leveraging the mobile small cell paradigm, in this paper we study an energy efficient Network-Coded Cooperative (NCC) scheme between mobile devices, which is an extension of the network coding (NC) based cooperative solution introduced in [3]. The performance of our NCC solution is studied in the context of Dense Networks (DNs) for mobile devices of different power supply capabilities and for a varying density of base stations and mobile devices, which is an important factor to take into account when cells are formed. We compare our solution with two baseline scenarios. The first scenario considers the case with no mobile small cells where mobile devices are connected to and served by a base station via a direct link. The second scenario corresponds to a baseline cooperative relaying solution where a mobile device acts as a relaying node forwarding data of other mobile devices to the base station without employing network-coding [4].

The remainder of this paper is organized as follows: The prior art is discussed in Section II and Section III introduces the energy efficient cooperation scheme based on network coding. Section IV presents simulation results for the energy efficiency and investigates its performance for different densities of BSs and mobile devices. Finally, conclusions are drawn in Section V.

## II. RELATED WORK

The combination of dense infrastructure networks and cooperative transmission between mobile devices has good potential for high efficiency. To this end, certain communications technology candidates are expected to play a fundamental role for the successful deployment of 5G systems [5]: cooperation, network coding, smart frontend, and small cells. Cooperative relaying has been introduced to improve performance and reliability of wireless networks [6], whilst network coding has been introduced to increase the throughput of wireless networks [7]. However, conventional data packet combining methods result in a higher sensitivity to error propagation [8]. Accordingly, cooperative relaying and network coding have complementary merits and limitations, and, in order to take advantage of their key benefits while overcoming their main limitations, a more innovative communication paradigm that is known as NCC communications has been proposed [9].

In general, cooperative relay techniques enhances the system performance mainly by means of improving the channel quality between the UE[1] (User Equipment) and the BS [10][11]. Cooperative relay techniques also allows to make use of diversity in the spatial and temporal dimension to mitigate the effects of fading and therefore to increase the reliability of radio links in wireless networks. In a dense wireless network, there are typically many relay nodes deployed. Determining which of these relay nodes should be selected for cooperation and how to cooperate with the selected relay nodes is a complicated problem [12]. Some schemes are proposed on this issue from a different viewpoint in [13].

On the other hand, network coding is a promising technique [7] to further increase the efficiency of cooperative transmission schemes. In brief, network coding allows intermediate nodes between the source and the destination to process and encode data packets from different inputs and to produce new output packets to forward [14]. Various NC schemes have been proposed for both wired and wireless networks, primarily for its ability to improve network throughput and robustness [15][16][17][18]. Some studies on energy savings by network coding have also been conducted. It has been demonstrated that NC can be an efficient way to improve the energy efficiency of a wireless sensor network compared to store-and-forward with traditional routing [19]. In [20] analogue NC is applied at relay nodes before forwarding signals to the destination nodes in multiple relay cooperative networks, targeting to minimize the energy consumption for a given transmission rate. In [21] NCC ARQ has been evaluated in a WLAN setting and energy efficiency improvements were observed. The energy efficiency improvements are even higher in [22] where an adaptive cooperative NC-based medium access control (MAC) protocol that exploits idle mobile devices as relays is proposed. In addition to minimizing the energy consumption of the mobile devices, different MAC protocol approaches combined with network-coded transmission also indicate promising performance in terms of network throughput and energy efficiency [23][24]. The practical feasibility of network coding processing in a UE has been demonstrated in [25]. Some research in the literature also shows that the performance of cooperative schemes largely depends on the locations of cooperation nodes. In [26], the region where cooperation is beneficial is described by an ellipse centred at the midpoint between the sender and the receiver.

---

[1] The terms UE and mobile device will be used interchangeably.

## III. NETWORK-CODED COOPERATIVE SCHEME FOR ENERGY EFFICIENCY

### A. Mobile Small Cells – Basic Concepts

In the context of this paper, the cooperative schemes form a Mobile Small Cell (MSC) consisting of a group of UEs, which aims at improving the energy efficiency by transmitting signals cooperatively. As depicted in Fig. 1, one of the group member UEs in the MSC is acting as the group head (GH). The term GH denotes a mobile device (or UE) that encodes and forwards data of other UEs within the wireless coverage of their MSC. In this paper we refer to the link between the GH and the BS as the fronthaul link since the processing of the network decoding and detection would be performed at the BS. In the use case scenario shown in Fig. 1, BSs are further connected to backhaul nodes or core network nodes via fixed or wireless connections.

The selection of the GH is one of the problems that will be guided by the evaluation in this paper. In particular, the relative locations of the MSC members and the BSs must be taken into consideration to achieve maximum throughput. However, since the GH will suffer from increased energy consumption, we study two different scenarios for the GH selection. As shown in Fig. 1, for the GH we may select:

1) a special mobile node which is not limited by battery constraints, for example corresponding to a mobile node mounted on vehicles such as busses and cars;

2) an ordinary UE with good link to the BS and preferably with no battery limitations.

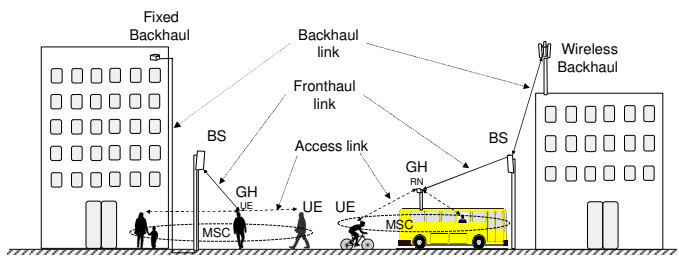

Figure 1. Use case scenarios of mobile small cells.

In this latter scenario the UE who acts as the GH of the MSC does not have any immediate benefit from the cooperation. However, if all UEs move randomly in the area of MSC it can be assumed that they all have equal probability to undertake GH role and hence all of them benefit from the cooperation in the longer term. For the first scenario we assume that the GH has an infinite energy supply, therefore the energy consumption of the GH can be ignored when calculating the energy efficiency. But the maximal transmit power of the GH is still restricted as an ordinary UE. For the second scenario, the energy consumption of all UEs including the GH is taken into account when studying the energy efficiency. In this paper the energy efficiency is defined as the ratio between the transmitted bits and the transmitted power.

Furthermore, it is assumed that mobile small cells, which are less involved than full-functioning BSs, provide basic functionalities including, encoding, relaying and network coding, along with transceiver functionalities (i.e. the ability to send and receive data). In addition, the role of GHs is to provide wireless front-haul, which constitutes the best available wireless link among all available wireless links, depending on network conditions and locations of the mobile devices.

### B. Network Coding Cooperative Scheme

In the NC cooperative scheme a pair of UEs in one MSC firstly send signals to the related MSC which then demodulates and decodes the two signals into two binary bit sequences. The MSC combines these two bit sequences to one bit sequence by network coding, thereafter the MSC encodes and modulates the combined bit sequence to one new signal and forwards it to the BS. This is done in a half-duplex mode, where the combined signal is forwarded in a separate time slot. Three links are involved in this procedure, we denote them as direct link, access link and fronthaul link. Direct link refers to the link between UE and BS, access link refers to the link between UE and MSC, while fronthaul link refers to the link between MSC and BS, as shown in Fig. 2. The channels of all links are modelled as quasi-static Rayleigh flat fading channels, and perfect channel knowledge is assumed. The channels could for example be efficiently estimated using support vector machines as described in [27]. Furthermore, this paper focuses on the uplink.

Assume mobile devices intend to transmit binary bit streams $u_i \in \{0,1\}, i = 1,2$ with subindex $i$ denoting UE1 and UE2. Let us suppose the binary bits after channel coding are $e_i \in \{0,1\}$, $i = 1,2$ where the channel coding scheme (e.g., turbo coding) follows a linear function $\Gamma(x)$, i.e.,

$$\Gamma(u_i) = e_i, i = 1,2. \qquad (1)$$

It is also assumed that the original binary information bits sent from UE1 and UE2 ($u_i$) are independently and uniformly distributed. Taking Binary Phase-Shift Keying (BPSK) modulation as an example, the modulation symbols are given by:

$$s_i = 1 - 2e_i, i = 1,2. \qquad (2)$$

The procedure of signal transmission is divided into three steps, described in the following:

*1) A pair of UEs transmit signals to GH via access link:* UE1 transmits signal $s_1$ and UE2 transmits signal $s_2$ to the GH (UE0 in Fig. 2) respectively. The received signals by UE0 are given by:

$$r_1^a = \sqrt{p_1} h_1^a s_1 + I_1^a + n_1^a, \qquad (3)$$

$$r_2^a = \sqrt{p_2} h_2^a s_2 + I_2^a + n_2^a, \qquad (4)$$

where $p_1$, $p_2$ are the transmitted signal powers of UE1 and UE2; $h_1^a$, $h_2^a$ are the channel gains of the access links; $n_1^a$, $n_2^a$ are noise contributions, and $I_1^a$, $I_2^a$ are the sum of interference generated from other UEs in the whole network.

Note that the signals sent by UE1 and UE2 can be also received by the BS (via direct link) due to the broadcast property of the wireless channel, thus:

$$r_1^d = \sqrt{p_1} h_1^d s_1 + I_1^d + n_1^d, \qquad (5)$$

$$r_2^d = \sqrt{p_2} h_2^d s_2 + I_2^d + n_2^d, \qquad (6)$$

where $h_1^d$, $h_2^d$ are the channels of the direct links between the UE1 and UE2 and the BS respectively; $n_1^d$, $n_2^d$ are the respective noise contributions; and $I_1^d$, $I_2^d$ are the respective sum of interference generated from other UEs in the whole network. The superscript '$a$' of variables denotes the access link and '$d$' the direct link. In this paper, noise and interference at the receivers are assumed to be Gaussian distributed.

The estimated signal $\tilde{s}_1$ and $\tilde{s}_2$ at UE0 can be derived from (3) and (4) as follows:

$$\tilde{s}_1 = \frac{r_1^a (h_1^a)^*}{\sqrt{p_1}\|h_1^a\|^2} = s_1 + \frac{(h_1^a)^*(I_1^a + n_1^a)}{\sqrt{p_1}\|h_1^a\|^2}, \qquad (7)$$

$$\tilde{s}_2 = \frac{r_2^a (h_2^a)^*}{\sqrt{p_2}\|h_2^a\|^2} = s_2 + \frac{(h_2^a)^*(I_2^a + n_2^a)}{\sqrt{p_2}\|h_2^a\|^2}, \qquad (8)$$

where $(h_1^a)^*$, $(h_2^a)^*$ are the complex conjugates of $h_1^a$, $h_2^a$. The power of the equivalent noise can be expressed as follows:

$$(\sigma_1^a)^2 = \frac{\|I_1^a\|^2 + \|n_1^a\|^2}{p_1 \|h_1^a\|^2}, \qquad (9)$$

$$(\sigma_2^a)^2 = \frac{\|I_2^a\|^2 + \|n_2^a\|^2}{p_2 \|h_2^a\|^2}. \qquad (10)$$

For BPSK, the corresponding bit error probabilities of $e_1$ and $e_2$ over the access link are formulated as follows:

$$P_e^{a_1} = \frac{1}{2} erfc\left(\frac{1}{\sqrt{2}\sigma_1^a}\right), \qquad (11)$$

$$P_e^{a_2} = \frac{1}{2} erfc\left(\frac{1}{\sqrt{2}\sigma_2^a}\right), \qquad (12)$$

where $erfc(x)$ is the complementary Gauss error function defined as follows:

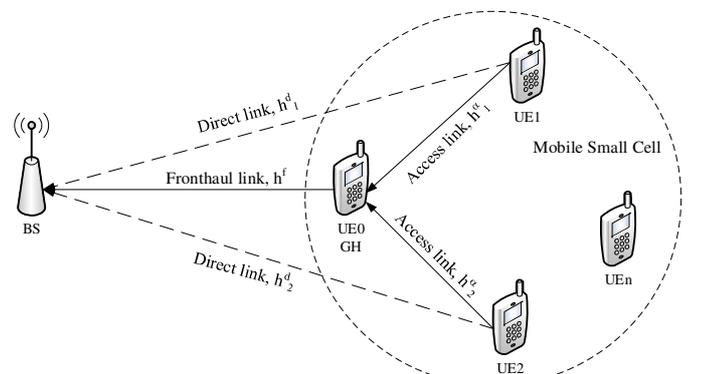

Figure 2. Mobile Small Cells and links.

TABLE I. Binary Bits and Modulated Signals Mapping

| Case | $e_1$ | $e_2$ | $e_1 \oplus e_2$ | $s_1$ | $s_2$ | $s'$ | $c$ |
|---|---|---|---|---|---|---|---|
| 1 | 0 | 0 | 0 | 1 | 1 | 1 | $c_1$ |
| 2 | 0 | 1 | 1 | 1 | -1 | -1 | $c_2$ |
| 3 | 1 | 0 | 1 | -1 | 1 | -1 | $c_3$ |
| 4 | 1 | 1 | 0 | -1 | -1 | 1 | $c_4$ |

$$erfc(x) = \frac{2}{\sqrt{\pi}} \int_x^\infty e^{-t^2} dt. \qquad (13)$$

*2) Fronthaul link transmission*: Assuming the UE0 demodulates and decodes $s_1$ and $s_2$ successfully, it can proceed with the original bit streams $u_1$ and $u_2$. A combined binary bit sequence $u' = u_1 \oplus u_2$ is generated by applying Exclusive OR (XOR) operation between $u_1$ and $u_2$ bit by bit.

Then, after applying channel coding to $u'$, i.e.

$$\Gamma(u') = \Gamma(u_1 \oplus u_2) = \Gamma(u_1) \oplus \Gamma(u_2) = e_1 \oplus e_2. \qquad (14)$$

UE0 modulates $\Gamma(u')$ to $s'$ and forwards it to the BS, thus $s'$ can be formulated as:

$$s' = 1 - 2e_1 \oplus e_2. \qquad (15)$$

The mapping relationship between binary bits $e_i$ and symbols for BPSK modulation, $s_i$ and $s'$, is shown in Table I. The vectors $c_i = (s_1, s_2, s')$, $i = 1,2,3,4$ are mapped to four points of a 3-dimensional constellation diagram $\mathcal{C}$, as shown in Fig. 3. The signal received at the BS from UE0 is given by:

$$r^f = \sqrt{p_0} h^f s' + I^f + n^f, \qquad (16)$$

where $p_0$ is the transmitted signal power of UE0; $h^f$ is the channel gain of the fronthaul link between UE0 and the BS; $n^f$ is the noise, and $I^f$ is the sum of interference generated from other UEs in the whole network. Again, the superscript '$f$' of variables denotes the fronthaul link, and the received noise and interference at the receivers are assumed to be Gaussian distributed.

Similarly, the equivalent noise power at the fronthaul link can be expressed as follows:

$$(\sigma^f)^2 = \frac{\|I^f\|^2 + \|n^f\|^2}{p_0 \|h^f\|^2}. \qquad (17)$$

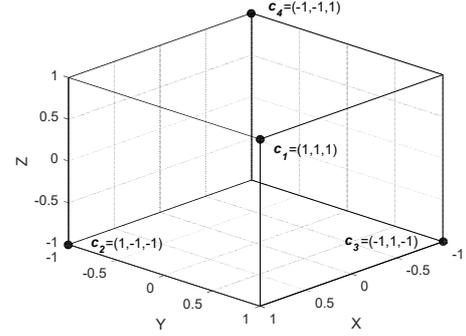

Figure 3. Network coding constellation diagram.

Thus the bit error probability of $s'$ at the fronthaul link is formulated as follows:

$$P_e^f = \frac{1}{2} erfc\left(\frac{1}{\sqrt{2}\sigma^f}\right). \qquad (18)$$

*3) Signal estimation at BS*: Let us take case 1 ($e_1 = 0$, $e_2 = 0$) as an example and later prove symmetry for the other cases. Then, (5) and (6) are converted to:

$$r_1^d = \sqrt{p_1} h_1^d + I_1^d + n_1^d, \qquad (19)$$

$$r_2^d = \sqrt{p_2} h_2^d + I_2^d + n_2^d. \qquad (20)$$

Let us define the estimated received signals at the BS $\tilde{r}_1^d$, and $\tilde{r}_2^d$, as:

$$\tilde{r}_1^d \triangleq 1 + \frac{(h_1^d)^*(I_1^d + n_1^d)}{\sqrt{p_1}\|h_1^d\|^2} = 1 + \tilde{n}_1, \qquad (21)$$

$$\tilde{r}_2^d \triangleq 1 + \frac{(h_2^d)^*(I_2^d + n_2^d)}{\sqrt{p_2}\|h_2^d\|^2} = 1 + \tilde{n}_2. \qquad (22)$$

Correspondingly we define a variable $\tilde{r}^f$ on the fronthaul link as follows:

$$\tilde{r}^f \triangleq 1 + \frac{(h^f)^*(I^f + n^f)}{\sqrt{p_0}\|h^f\|^2} = 1 + \tilde{n}_3. \qquad (23)$$

With the assumption that $\tilde{n}_1$, $\tilde{n}_2$, $\tilde{n}_3$ are independent and normally distributed with zero mean and with variances equal to $\tilde{\sigma}_1^2$, $\tilde{\sigma}_2^2$, $\tilde{\sigma}_3^2$ respectively, the joint probability distribution function of $\tilde{n}_1$, $\tilde{n}_2$ and $\tilde{n}_3$ is given by:

$$p(x,y,z) = \frac{1}{(\sqrt{2\pi})^3 \tilde{\sigma}_1 \tilde{\sigma}_2 \tilde{\sigma}_3} e^{-\frac{x^2}{2\tilde{\sigma}_1^2} - \frac{y^2}{2\tilde{\sigma}_2^2} - \frac{z^2}{2\tilde{\sigma}_3^2}}, \quad (24)$$

Based on the above definitions vector $\tilde{r}$ can be defined as:

$$\tilde{r} \triangleq (\tilde{r}_1^d, \tilde{r}_2^d, \tilde{r}^f), \quad (25)$$

which corresponds to a vector in the 3-dimensional space, depicted in Fig. 3. Since there are four constellation points in diagram $\mathcal{C}$, the transmitted symbols $s_i$ will be estimated according to the Euclidian distance between $\tilde{r}$ and the constellation points $c_i$, henceforth denoted as $d_i = \|\tilde{r} - c_i\|^2$. Bearing in mind Table I and the constellation diagram in Fig. 3, the correct detection of, say, transmitted symbol $s_1 = 1$ (equiv. $e_1 = 0$) requires either of the following two events to occur: event **A**, whereby distance $d_1$ is smaller than distances $d_3$ and $d_4$; or event **B**, whereby distance $d_2$ is smaller than distances $d_3$ and $d_4$. Event **B** corresponds to the case where $u_2$ is incorrectly decoded. Therefore, and by virtue of the addition rule, the correct detection probability $P_c$ for the transmitted symbol $s_1 = 1$ ($e_1 = 0$) can be formulated as:

$$P_c = \boldsymbol{P}(\boldsymbol{A} \cup \boldsymbol{B}) = \boldsymbol{A} + \boldsymbol{B} - \boldsymbol{A} \cap \boldsymbol{B}, \quad (26)$$

with

$$\boldsymbol{A} = P\{d_1 \leq d_3, d_1 \leq d_4\} = \iiint_{\Lambda_A} p(x,y,z) dz dy dx, \quad (27)$$

where

$$\Lambda_A = \{(x,y,z) | x+y+2 \geq 0, x+z+2 \geq 0, x,y,z \in \mathbb{R}\};$$

and with

$$\boldsymbol{B} = P\{d_2 \leq d_3, d_2 \leq d_4\} = \iiint_{\Lambda_B} p(x,y,z) dz dy dx, \quad (28)$$

where

$$\Lambda_B = \{(x,y,z) | x-y \geq 0, x-z \geq 0, x,y,z \in \mathbb{R}\}.$$

The calculation of correct detection probability for the rest of cases ($e_1 = 1$, $e_2 = 0$ and $e_2 = 1$, and in general for any $e_i$) is exactly the same as in (26)-(28) since the distribution of the four constellation points is symmetric and we assume that the original bits are independent identically distributed.

For convenience, we generalize the notation $P_c$ to reflect all different cases and redefine $P_c$ in (26) as:

$$P_c(s_i|s_1, s_2, s') \overset{\text{def}}{=} P_c, i = 1,2, \quad (29)$$

where $P_c(s_i|s_1, s_2, s')$ should be read as the correct detection probability of symbol $s_i$ given that symbols $s_1$ and $s_2$ where transmitted by UEs 1 and 2, and where NC operation produced transmitted symbol $s'$.

At this point the following observation can be made. The correct symbol detection probability at the BS partially depends, as reflected in (29), on the transmitted symbol from the GH over the fronthaul $s'$. This means that, even if the transmitted symbol $s'$ contained errors, a correct detection could still be obtained at the BS (e.g. by virtue of the direct transmission paths and/or the XOR operation via NC). The ability by which the GH can realize if the transmitted symbol $s'$ contains errors (e.g. through Cyclic Redundancy Check, CRC) allows us to characterize the GH (and hence the fronthaul) according to different implementation capabilities, two of which are addressed in this paper, and presented next:

*a) GH with no error detection mechanisms (no CRC)*

In this implementation, no cyclic redundancy check (CRC) is available at the GH, thus acting as an uninvolved network node with low complexity. Since it requires a change to the physical layer processing it would correspond to realization of the fronthaul link by a dedicated relay node. For this case, the calculation of the NC-based correct detection probability of symbol $s_1 = 1$ provided the transmission of symbols ($e_1 = 0$, $e_2 = 0$), is a combination of two component probabilities corresponding to the following two events:

Event I ($\boldsymbol{E_I}$): The correct detection probability of transmission $s_1 = 1$ considering the case when GH transmits correct network coded signals to the BS. For the considered example ($e_1 = 0$, $e_2 = 0$) this means, $s' = 1$ is transmitted by the GH and the received signal over the fronthaul is $\tilde{r}^f = 1 + \tilde{n}_3$. The probability for the $\boldsymbol{E_I}$ event is given by:

$$\mathbf{P}(E_I) = [(1 - P_e^{a_1})(1 - P_e^{a_2}) + P_e^{a_1} P_e^{a_2}] \cdot P_c(s_1|s_1, s_2, s' = 1) \quad (30)$$

where the first factor within brackets is the probability that $s' = 1$ is transmitted, and the second factor is the probability of correctly detecting transmission $s_1 = 1$ at the BS given by (29).

Event II ($E_{II}$): The correct detection probability of transmission $s_1 = 1$ considering the case when GH transmits incorrect network coded signals to the BS. For the considered example ($e_1 = 0, e_2 = 0$) this means, $s' = -1$ is transmitted by the GH and the received signal over the fronthaul is $\tilde{r}^f = -1 + \tilde{n}_3$. The probability for this event, $E_{II}$, is given by:

$$\mathbf{P}(E_{II}) = [(1 - P_e^{a_1}) P_e^{a_2} + P_e^{a_1}(1 - P_e^{a_2})] \cdot P_c(s_1|s_1, s_2, s' = -1) \quad (31)$$

where the first factor within brackets is the probability that $s' = -1$ is transmitted, and the second factor is the probability of correctly detecting transmission $s_1 = 1$ at the BS given by (29).

Then, the overall correct detection probability considering the no error detection mechanism at the BS (no CRC) is given as:

$$P_c^1 = \mathbf{P}(E_I) + \mathbf{P}(E_{II}) \quad (32)$$

where $\mathbf{P}(E_I)$ and $\mathbf{P}(E_{II})$ are given by (30) and (31) respectively.

*b) GH with error detection mechanisms (CRC)*

A realisation of the fronthaul link with CRC at the GH requires less changes to the normal UE processing. This case corresponds to the expected realisation of the backhaul link provided between a UE, acting as GH, and a base station. While the correct detection probability of a fronthaul link with no CRC at the GH can be calculated by (32), the calculation of the correct detection probability of transmitted symbol $s_1 = 1$, provided the transmission of symbols ($e_1 = 0, e_2 = 0$), over fronthaul link with CRC at the GH is obtained by the following expression:

$$P_c^1 = [(1 - P_e^{a_1})(1 - P_e^{a_2})] \cdot P_c(s_1|s_1, s_2, s' = 1), \quad (33)$$

where the first factor is the probability that the received symbols at the GH are both correct.

Based on the correct detection probability in (32) and (33) the miss-detection probability of transmitted symbol $s_1 = 1$ with ($e_1 = 0, e_2 = 0$) is calculated as follows:

$$P_e^{nc_1} = 1 - P_c^1, \quad (34)$$

The bit error probabilities $P_e^{nc_2}$ of case 2 ($e_1 = 0, e_2 = 1$), $P_e^{nc_3}$ of case 3 ($e_1 = 1, e_2 = 0$), and $P_e^{nc_4}$ case 4 ($e_1 = 1, e_2 = 1$) can be calculated in the same way as above. Assuming that all four cases are equally probable, the average miss-detection probability is given by:

$$\bar{P}_e^{nc} = \frac{(P_e^{nc_1} + P_e^{nc_2} + P_e^{nc_3} + P_e^{nc_4})}{4} = P_e^{nc_1}. \quad (35)$$

In other words, due to the symmetry of the four cases the overall bit error probability $\bar{P}_e$ is equal to the first case.

*C. Forwarding Scheme*

Similarly to the network coding scheme, the bit error probability of the forwarding scheme, $P_e^{fw}$, depends on the implementation of the fronthaul link and the functional capabilities of the selected GH. If there is no CRC at the GH, the bit error probability $P_e^{fw}$ is given by

$$P_e^{fw} = P_e^a + P_e^f - 2 P_e^a P_e^f, \quad (36)$$

where $P_e^a$ is the bit error probability at the access link, and $P_e^f$ the bit error probability at the fronthaul link. While, if there is CRC at GH then the bit error probability is obtained by

$$P_e^{fw} = P_e^a + P_e^f - P_e^a P_e^f. \quad (37)$$

It can be seen that the case with no CRC is expected to perform better in general since the BS may decode an erroneous signal only when either the access link or the fronthaul link is erroneous in an exclusive manner. The BS will be able to correctly decode the signal in the case when the signals received by both links are erroneous. While in the case of CRC the forwarding is performed only when the signal of the access link is correctly received by the GH followed by a correctly received signal by the BS over the fronthaul.

*D. Transmit power optimisation*

The transmission power of each UE is optimized to minimize the total energy consumption under a constraint on the maximum bit error probability. The resulting optimization problem can be formulated as follows:

$$\min f(p_i) = \sum_{i=1}^{N} p_i \quad (38)$$

$$s.t.\ P_{e_i} \leq \vartheta_e,\ P_{min} < p_i \leq P_{max}, i = 1,2,\dots,N,$$

where $\vartheta_e$ is the threshold of maximum bit error probability, $P_{e_i}$ is the bit error probability of $UE_i$, $P_{max}$ and $P_{min}$ are the maximal and minimal transmission powers of UE respectively, and $N$ is the number of UEs.

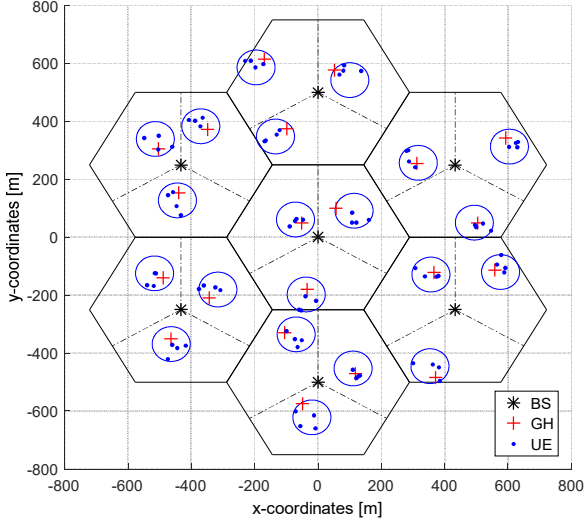

Figure 4. Cellular deployment for performance evaluation.

In this paper, the optimization problem (38) is solved by the interior point algorithm [28], which follows a barrier approach that employs sequential quadratic programming (SQP) and trust regions to solve the sub-problem occurring in each iteration. The SQP method solves a sequence of optimization sub-problems, each of which optimizes a quadratic model of the utility function to a linearization of the constraints. Once the optimal power configurations of UEs are identified, the capacity of each link is approximated by the Shannon equation.

## IV. SIMULATION RESULTS

### A. System model

For the performance evaluation a network consisting of a hexagonal grid of 7 base station sites is considered. Each cell is divided into 3 equally sized sectors by means of directional antennas, as illustrated in Fig. 4. For the uplink (UL) transmission power control is applied so that a UE that has a large distance to the serving BS (e.g., the UE located at the cell edge area) increases the transmit power to compensate the path loss of the wireless channel.

The configuration of simulation parameters, which are listed in Table II, is mainly based on the 3rd Generation Partnership Project (3GPP) spatial channel model [30]. We consider an urban microcell environment with BS antennae situated at rooftop height and a non-light of sight (NLOS) pathloss model at 2 GHz based on the COST 231 Walfish-Ikegami street canyon model [31].

In the simulations the GH is chosen as the UE who has the best channel quality towards a BS among all the members within one MSC. In the case where the GH is not battery constrained, it transmits signals at a maximal power allowed for a UE, while in the case where GH is constrained by the battery, the transmit power of GH is optimized in the objective function.

TABLE II. CONFIGURATION OF SIMULATION PARAMETERS

| Parameter | Values and Assumptions |
|---|---|
| Inter-Site Distance (ISD) | 500m |
| Carrier Frequency/Bandwidth | 2GHz/10MHz |
| UE/MSC maximum transmit power | $P_{max}$ = 250mW |
| UE/MSC minimum transmit power | $P_{min}$ = 0.1µW (cf. [29]) |
| UE distance in one group | $d_{min}$ = 35m, $d_{max}$ = 100m |
| UE power control for DT (Fractional power control) | $P_0$ = -110dBm, α = 1.0 (alpha) |
| Antenna pattern (horizontal) (For 3-sector cell sites with fixed antenna patterns) | $A(\theta) = -\min\left[12\left(\frac{\theta}{\theta_{3dB}}\right)^2, A_m\right]$ $\theta_{3dB} = 70^0, A_m = 20\text{dB}$ |
| Number of paths (K) and sub-paths per-path (M) | K = 6; M = 20; (cf. [30]) |
| Lognormal shadowing standard deviation | NLOS: 10dB |
| Mean Angle Spread (AS) at BS | $E(\sigma_{AS,\ BS})$ = 19° |
| Mean AS at UE[a] | $E(\sigma_{AS,\ UE})$ = 68° |
| Per-path AS at UE (fixed) | 35° |
| Per-path AS at BS (Fixed) | 5° |
| Mean total RMS Delay Spread | $E(\sigma_{DS})$=0.251µs |
| Distribution for path delays | U(0, 1.2µs) |
| Pathloss Model | 34.5 + 38log$_{10}$(d), d in m |
| Minimal throughput threshold per UE | 1 Mbps |

[a] UE = Mobile Station (MS) [30]

## B. Comparison with Direct Transmission

The traditional direct transmit (DT) scheme in which all the UEs transmit signals to the BS directly is used as a reference for the evaluation of the NC scheme. The transmit power of the UE in the DT scheme is governed by the UL power control used in the Long-Term Evolution (LTE) system [32], i.e. $P = P_0 + \alpha PL$, where $PL$ is the path loss and we take $P_0 = -110$dBm, $\alpha = 1$ in the simulation. The results are also compared to the forwarding (FW) scheme where each UE firstly transmits signals to the GH which thereafter forwards the signals to the BS. This corresponds to a scenario where GH acts as a relaying node of the MSC. In order to make the results comparable, the threshold maximum bit error probability $\vartheta_e$ is that of the DT. Additionally, the influence of fast fading is not included and it is assumed that the users always have data to transmit in their buffers.

Fig. 5 compares the performance of the DT, FW and NC schemes in terms of average throughput. Regarding the FW and NC scheme, the suffix 'WI'/'WO' in legend denotes the case that the MSC is assumed with or without good energy supply respectively, e.g., FW_WI represents the forwarding scheme assuming GH with continuous power supply, while NC_WO represents the network coding scheme assuming a battery powered GH and therefore its transmit power is part of the optimization. In the context of this paper, a GH without battery limitation is defined as one that has a continuous power supply or a large remaining energy when the expected energy consumption of the planned usage until the next charging is subtracted.

Fig. 5(a) illustrates the average throughput of the three schemes in the case of a GH with no CRC functionality. Overall, the NC scheme performs better than the FW scheme regardless of the power limitations of the GH. The gains made by FW and NC schemes over the DT scheme in terms of average UE throughput are quantified in Fig. 5(b). Both FW and NC schemes outperform the DT scheme. The largest gain is achieved by the NC which reaches up to 66.3% when a GH has no battery limitations. On opposite end, the performance of FW scheme assuming GH with battery limitations is quite close to the performance of the DT scheme. Finally, and as expected, the demonstrated gains of the FW and DT, when a GH with CRC functionality and no battery limitations is employed, are slightly lower and reach up to 45.4% and 62.5% for the FW and the NC scheme respectively.

Fig. 6 compares the performance of the DT, FW and NC schemes in terms of the average UE energy efficiency measured in Mbps/Watt. As is shown in Fig. 6(a), the schemes assuming GH without energy constraint achieve better performance than the schemes assuming GH with battery limitations for both FW and NC schemes. The reason is that a higher transmit power can be set if GH does not have to minimize the power due to the battery limitation.

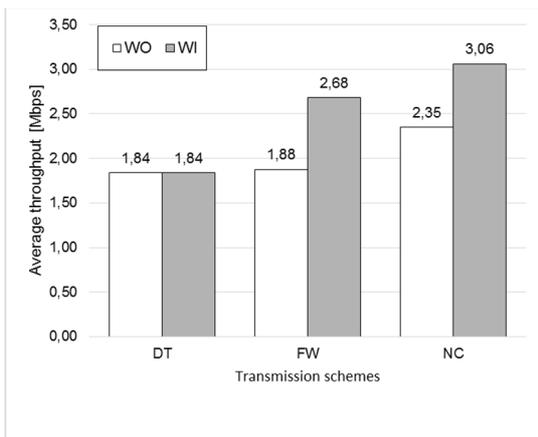

(a) Average throughput of transmission schemes with/without battery limitation.

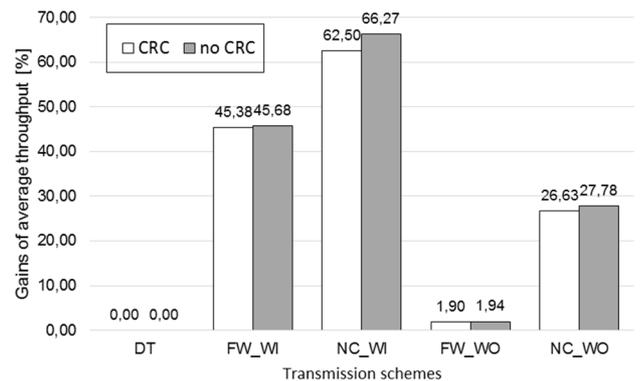

(b) Comparison of throughput gains with/without CRC at GH relative to DT scheme

Figure 5. Comparison of the throughput performance of the different transmission schemes

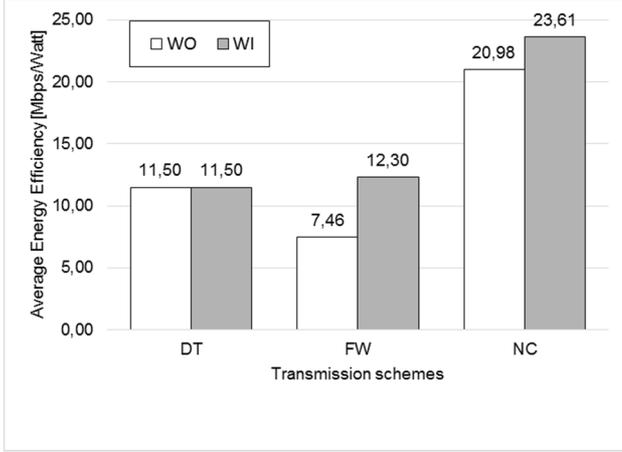
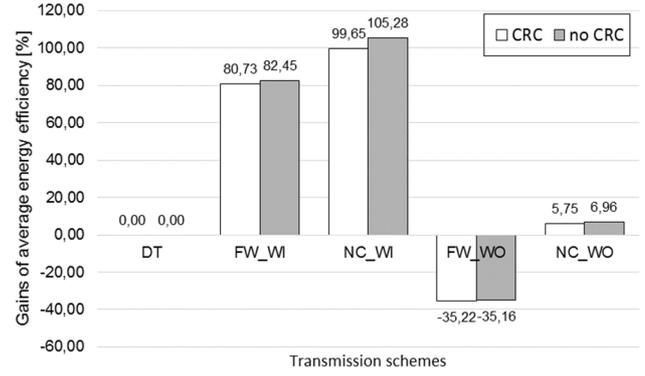

(a) Average energy efficiency of transmission schemes with/without battery limitation.

(b) Comparison of energy efficiency gains with/without CRC at GH relative to the DT scheme

Figure 6. Comparison of energy efficiency performance of the different transmission schemes

Fig. 6(a) also demonstrates that NC schemes overall outperforms their corresponding FW schemes of the same battery assumption about GH, i.e., NC_WI outperforms FW_WI, while NC_WO outperforms FW_WO. The gains made by FW and NC schemes over the DT scheme in terms of average energy efficiency are quantified in Fig. 6(b), where it is shown that both NC schemes perform better than the DT scheme. Furthermore, in the case of a GH with no CRC functionality, the largest gain is achieved by NC_WI scheme, which more than doubles (105.3%) the energy efficiency. Even for the NC_WO scheme an approximate gain of 6.9% gain is observed. While, for the FW scheme, the gains reaches 82.5% when the GH is not limited by the battery. However, for the case of GH with battery limitation a degradation of about 35% compared with the DT scheme is observed. Finally, for a GH with CRC functionality and continuous power supply the corresponding gains of the FW and the NC schemes are 80.7% and 99.7% respectively.

## C. Energy Efficiency Dependence on the Density of Cooperation Nodes

The performance of the network coding scheme largely depends on the density and the locations of cooperation nodes and the density of the base stations. In this section we investigate the scheme with the goal of characterizing scenarios where it is beneficial for devices to cooperate. This can be used to form MSCs of cooperating devices. The formation of the MSCs is performed when there is a gain from cooperation and avoided when the gains do not justify the overhead. Since the advantage of network coding depends on the relation of at least four nodes, the relations of a few typical characteristics are investigated by means of numerical simulation for two different scenarios. In the first scenario the density of the base stations increases while the user density is fixed. In the second simulation scenario the base station density is kept fixed while the user density increases.

Fig. 7 depicts the energy efficiency in Mb/Joule of the DT, FW and NC schemes for different base station densities and a user density of 42 UEs/km$^2$. The user density implies that the locations of the UEs and MSC are on average 100 meters apart which also characterizes the average distance of the access links. In this scenario, the NC_WI scheme achieves better performance than the other schemes when the density of the base stations falls below 82 BSs/km$^2$ implying a distance of the direct link larger than 150m on average. The performance of NC_WO is quite close to that of baseline up to a base station density of approximately 40 BSs/km$^2$. Within this range the NC scheme assuming MSC without battery limitation performs significantly better than all the other schemes, while the FW_WO scheme performs worse than all the other schemes at all observed cases. Overall, the cooperative schemes perform worse than the DT scheme when the location of MSC is very close to BS, which is reflected when the base station density receives its highest value of 265 BSs/km$^2$. Furthermore, in such a dense UDN, the distance of the fronthaul link and the direct link reaches 40m and 125m respectively.

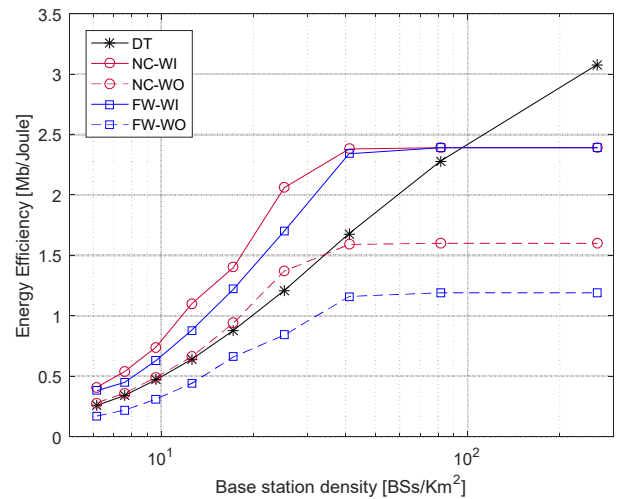

Figure 7. Energy efficiency comparison of DT, FW and NC schemes for different base station densities.

In general, when the difference between the distance of access link and that of the direct link is small, the gains in bit error probability produced by the cooperative schemes are relatively minor, and do not compensate for the additional bandwidth consumed in the cooperative schemes. The performance gains of the NC scheme over the forwarding disappears when the distance of the fronthaul link exceeds 200m, which is the case when the base station density is below 8 BSs/km$^2$. From the results in Fig. 7, it can be concluded that it is not worthwhile to form an MSC with a GH that is too close to the BS.

In Fig. 8, it is shown the energy efficiency of the baseline and cooperative schemes in the case where the base station density is fixed while the UE density varies. In this simulation scenario the base station density equals 42 BSs/km$^2$ resulting in a fronthaul distance of 100m on average. A decrease in the UE density implies an increase of the distance between the UEs and the GH of the MSC. As a consequence, the NC_WI scheme achieves better performance than the other schemes when the UE density is up to 512 UEs/km$^2$ implying an average distance of the UE access link that exceeds 75m.

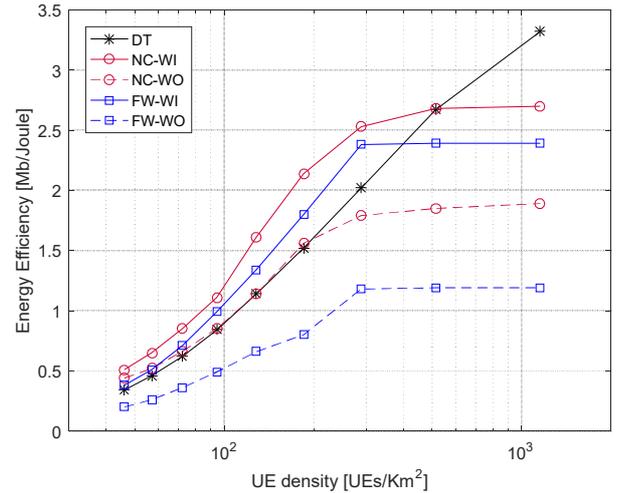

Figure 8. Energy efficiency comparison of DT, FW and NC schemes for different UE densities.

### D. Performance Evaluation Remarks

In summary, the network coding and forwarding schemes achieves their best performance when the fronthaul link has a length comparable with the access link. Although previous studies have shown throughput gains also when a normal battery powered UE acts as a GH [3], there are no substantial energy efficiency gains in such a scenario. The cooperative schemes are therefore mainly of interest in a deployment scenario where the network is densified with GHs employed by devices that are mounted on locations with electricity supply but no fixed backhaul, such as bus stops, busses, cars and other public transportation vehicles. Using mobile devices has the advantage that the GHs do not need to be battery powered and that they are moving along with the users.

The network coding scheme has the widest range of relative densities and locations that provide energy efficiency gains, and is therefore preferable when it can be applied. It may also be extended to encoding of transmissions from more than two users. However, from the results it has been demonstrated that a network coding scheme is limited. First, it is only applicable when there are multiple users whose transmissions can be relayed together. For this reason, the forwarding scheme is an important complement which is also easier to implement with limited changes to current technology by building on LTE relay node or side link specifications. Second, employing NC, or any other cooperation scheme, implies a higher energy cost in the case of dense networks. In deployment scenarios that aim at supporting indoor users or high-speed users and highly dense deployed vehicles in urban areas, the inter-site distances, e.g., of the roadside units, range between 20m (indoor) and 50m (outdoor) [33]. When the energy efficiency is the only consideration, these deployments would benefit by employing a mere DT scheme rather than a cooperative scheme with high overhead, such as, NCC.

## V. Implementation Aspects

In this paper we utilise an optimization approach to derive an upper bound on the energy efficiency performance for different cooperative transmission schemes and implementations of MSCs. In this section some practical implementation aspects of these schemes are discussed including power control, MSC formation and selection of GH, as well as observations on UE/BS density and frequency utilisation.

In particular, we consider a UL transmit power optimization for the UEs that use the network coding scheme, although the discussion can be generalised for any cooperative scheme. The intention is that the BS would calculate the optimal transmission power for each UE according to (38) and use the UL transmission power control procedures to regulate the transmit power of the UEs. In a practical algorithm, the measured block-error probability and the channel qualities could be used directly to achieve the target performance with low complexity. In one example solution, by leveraging 3GPP LTE power control mechanisms in 5G, a closed loop power control mechanism which is under the control of the BS and triggered by the channel dynamics can be considered. Since 3GPP only specifies the UE behaviour it is feasible to implement power control algorithms in the BS that controls the UL transmit power with different targets. It is therefore only necessary to recalculate the transmission power when the SINR changes significantly. How often that happens depends on the user mobility, but by limiting the use of cooperative schemes to users that are relatively stationary it can be tens to hundreds of transmission time intervals (TTIs) depending on the environment.

Furthermore, knowledge of UE positions may also allow for a BS to determine which UE should belong to the MSC and which UE should act as the GH. Such a centralised solution has been proposed in [34] where the BS groups UEs based on their mobility and determines which UE within the group should act as the GH. For the selection of the GH the BS may request additional information among the UEs in the group such as battery supply, i.e., whether they are battery powered or they have access to continuous energy supply, and UE capabilities, i.e., whether they can operate as UE or fully functional relay nodes. A distributed solution would require a discovery mechanism that facilitates UEs to identify each other and exchange information necessary for building the MSCs. By means of a discovery protocol based on UL beacons, UEs may exchange information such as received power, fronthaul link channel state and battery level indicator. The central UE with the best fronthaul link and a high battery level can be selected as the GH which may in addition control the transmit power of the UEs in the MSC.

Certainly, such a discovery mechanism can be extended to implement a hybrid solution which also includes the discovery of BSs that perform power control and GH selection. Depending on the UE density and the density of its neighbour BSs, a BS may also determine the cooperation transmission scheme to be employed by MSCs and carried over their access links and fronthaul links. This is particularly useful in a dynamic radio environment where interference may significantly fluctuate due to a varying number of mobile devices and activated/deactivated BSs operating power saving mode during a day, implying diurnal density fluctuations.

Another characteristic of 5G networks is the wider range of carrier frequency bands that will be supported. In particular, for higher frequencies it is necessary to consider the directionality of the transmission. Wireless network coding schemes in general rely on the broadcast nature of the radio transmission to achieve the benefits of cooperation. With beam formed transmission this does no longer come for free, and the benefit of network coding is diminished. The transmitted power would need to be divided into multiple directional transmissions, which would reduce the received power in each beam.

## VI. CONCLUSIONS

In this paper an energy efficient cooperation scheme based on network coding to improve the energy efficiency for users in a mobile small cell is studied. A mobile small cell consists of a number of users in a certain contiguous area. The network coding scheme is implemented by cooperatively transmitting signals among users in one mobile small cell. One of the users is selected as the group head to coordinate the cooperation among the users within the mobile small cell. The performance of the network coding scheme is evaluated by means of simulations in a dense urban cellular network by two scenarios: 1) the group head is equipped with infinite power supply, e.g., a laptop or femtocell with LTE fronthaul without CRC and fixed connection to the main power; 2) the group head is an ordinary mobile device, e.g., a smart phone with portable battery. The numerical simulation results show that both cooperative schemes, i.e., the network coding scheme and the forwarding scheme, outperform the direct transmit scheme for low and moderate base station density scenarios. In high-density scenarios direct transmission is the preferred choice when energy efficiency is the only consideration. In the low-density scenarios, where the cooperative schemes are of preference, the gains are mainly due to the high transmission power at the group head, therefore the energy efficiency is worse than direct transmission when the transmission power of the GH is taken into account. To achieve significant gains the MSC should therefore be formed around a GH with low sensitivity to energy consumption. Since the performance of the cooperation schemes largely depends on the locations of cooperating nodes these schemes shall only be applied for certain favourable densities of base stations and cooperating nodes. As a rule of thumb, a GH should be located between the cooperating UEs and the BS at densities where the access link has a comparable distance with the fronthaul link.

## ACKNOWLEDGMENT

This research work has received funding from the European Union's H2020 research and innovation program under grant agreement H2020-MCSA-ITN- 2016-SECRET 722424 [35].